\documentclass[conference]{IEEEtran}
\IEEEoverridecommandlockouts
\usepackage{cite}
\usepackage{amsmath,amssymb,amsfonts}
\usepackage{algorithmic}
\usepackage{graphicx}
\usepackage{textcomp}
\usepackage{xcolor}
\usepackage{upgreek}
\usepackage{mathtools}

\def\BibTeX{{\rm B\kern-.05em{\sc i\kern-.025em b}\kern-.08em
    T\kern-.1667em\lower.7ex\hbox{E}\kern-.125emX}}

\usepackage{multirow}
\usepackage{makecell}
\usepackage{colortbl}
\newcolumntype{C}[1]{>{\centering\arraybackslash}p{#1}}
\newcolumntype{M}[1]{>{\centering\arraybackslash}m{#1}}

\usepackage{booktabs,siunitx}
\usepackage{tabularx}
\usepackage[none]{hyphenat}%%%%
\newcommand{\cmark}{\checkmark}%
\newcommand{\xmark}{\small$\boldsymbol\times$}%

\begin{document}

\title{Real-time Speech Enhancement and Separation with a Unified Deep Neural Network for Single/Dual Talker Scenarios \\
\thanks{This work was supported by the National Institute on Deafness and Other Communication Disorders (NIDCD) of the National Institutes of Health (NIH) under Award 5R01DC015430-05. The content is solely the responsibility of the authors and does not necessarily represent the official views of the NIH.}
}

% \author{\IEEEauthorblockN{1\textsuperscript{st} Kashyap Patel}
% \IEEEauthorblockA{\textit{Electrical and Computer Engineering} \\
% \textit{University of Texas at Dallas}\\
% Richardson, USA \\
% patelkashyap@utdallas.edu}
% \and
% \IEEEauthorblockN{2\textsuperscript{nd} Anton Kovalyov}
% \IEEEauthorblockA{\textit{Electrical and Computer Engineering} \\
% \textit{University of Texas at Dallas}\\
% Richardson, USA \\
% anton.kovalyov@utdallas.edu}
% \and
% \IEEEauthorblockN{3\textsuperscript{rd} Issa Panahi}
% \IEEEauthorblockA{\textit{Electrical and Computer Engineering} \\ 
% \textit{University of Texas at Dallas}\\
% Richardson, USA \\
% imp015000@utdallas.edu}
% }

\author{\IEEEauthorblockN{Kashyap Patel\IEEEauthorrefmark{1},
Anton Kovalyov, and Issa Panahi}
\IEEEauthorblockA{Electrical and Computer Engineering,
University of Texas at Dallas\\
Richardson, USA\\
Email: \IEEEauthorrefmark{1}patelkashyap@utdallas.edu}}

\maketitle

\begin{abstract}
This paper introduces a practical approach for leveraging a real-time deep learning model to alternate between speech enhancement and joint speech enhancement and separation depending on whether the input mixture contains one or two active speakers. Scale-invariant signal-to-distortion ratio (SI-SDR) has shown to be a highly effective training measure in time-domain speech separation. However, the SI-SDR metric is ill-defined for zero-energy target signals, which is a problem when training a speech separation model using utterances with varying numbers of talkers. Unlike existing solutions that focus on modifying the loss function to accommodate zero-energy target signals, the proposed approach circumvents this problem by training the model to extract speech on both its output channels regardless if the input is a single or dual-talker mixture. A lightweight speaker overlap detection (SOD) module is also introduced to differentiate between single and dual-talker segments in real time. The proposed module takes advantage of the new formulation by operating directly on the separated masks, given by the separation model, instead of the original mixture, thus effectively simplifying the detection task. Experimental results show that the proposed training approach outperforms existing solutions, and the SOD module exhibits high accuracy.\par
\end{abstract}


\section{Introduction}\label{sec:intro}
In recent years, a lot of research has been done on deep learning (DL)-based speech enhancement (SE) and speech separation (SS) methods. Applications include automatic speech recognition (ASR) and hearing aids \cite{kovalyov2023dfsnet, luo2018tasnet, bhat2019real}. In the real world, speech signals are often a reverberant mixture of one or more speech sources and noise. When the number of speech sources is not known a priori, it can be hard for a real-time system to determine whether only SE, in the case of a single talker, or a combination of SE and SS, in the case of multiple overlapping talkers, should be applied. In practice, the number of overlapping talkers is rarely more than two \cite{css}. Hence, for the purpose of this work, we consider only single and dual-talker scenarios.\par

Numerous techniques have been suggested to overcome the challenges of a practical system when dealing with different numbers of concurrently active talkers \cite{count_separate, nachmani2020voice, zhu2021multi, takahashi2019recursive, jiang2020speaker}. One approach is to train multiple networks for different speaker counts \cite{count_separate, nachmani2020voice}. The speaker count is then estimated by a separate module and the appropriate SE or SS model is selected to extract the speech signal/s. In a similar but more efficient approach, a noncausal neural network architecture was proposed that utilizes a shared encoder and separator but a different decoder for each speaker count \cite{zhu2021multi}, resulting in considerably less training parameters. Although the technique of speaker counting followed by selecting the appropriate SE or SS module was shown to work well in noncausal settings, it is unclear how it can be efficiently applied to latency-demanding applications.\par

% Recent research has also focused on incorporating silent targets in the loss functions during training of the SS model to address the problem of varying number of talkers in the mixture.     

In a more practical approach, a single-talker signal can be modeled as dual-talker by setting the second signal to zero energy, i.e., silent speech. With this formulation, an SS model can be trained to jointly perform SS and SE in either single or dual talker scenarios. However, time-domain SS models are typically trained using a signal-to-distortion ratio (SDR)-based loss function, which is ill-defined for zero-energy target signals. Although modifications to SDR-based loss functions have been proposed to handle zero-energy target signals \cite{von2022sa, luo2020separating}, they generally come at the cost of somewhat degraded performance in terms of the original SDR metric.\par

Motivated by the above observations, this study proposes a simple and efficient approach for training a DL-based SS model to handle both single and dual-talker scenarios. Given an input mixture, the proposed approach consists in training a model to output two channels. In a dual-talker scenario, these channels correspond to the two separated and enhanced speech signals, whereas in a single-talker case, both channels correspond to the same output, the enhanced speech. This approach allows leveraging the standard permutation invariant training (PIT) \cite{yu2017permutation} with an SDR-based loss function without requiring any modifications. Additionally, taking advantage of the new formulation, a lightweight speaker overlap detection (SOD) post-processing module is introduced for detecting dual-talker instances in real time. This module simplifies the detection task by operating directly on the separated masks, estimated by the SS model, instead of the original mixture signal. The proposed methodology was tested on the recently introduced UX-Net model \cite{patel2022ux} for causal, low-latency SS, revealing improved performance over methods that reformulate the SDR measure. The proposed SOD module is also shown to attain high accuracy.\par

\section{Problem Formulation} \label{sec:problem}
Let us consider a scenario with one or two active speech sources in a noisy and reverberant environment. The time-domain signal captured by the microphone is modeled by
\begin{equation} \label{eq:input}
    \mathbf{y} = \mathbf{s}_1 + \mathbf{s}_2 + \mathbf{v}
\end{equation}
where $\mathbf{s}_1$ and $\mathbf{s}_2$ are the clean reverberant speech signals of the two sources, and $\mathbf{v}$ is background noise. In a dual-talker scenario, we wish to extract both $\mathbf{s}_1$ and $\mathbf{s}_2$ from $\mathbf{y}$ (SS task), whereas, in a single talker scenario, $\mathbf{s}_2$ is assumed to be a zero-energy signal, and we wish to extract only $\mathbf{s}_1$ from $\mathbf{y}$ (SE task).\par

Let $S = \{\mathbf{s}_1, \mathbf{s}_2\}$ and $\hat{S} = \{\hat{\mathbf{s}}_1, \hat{\mathbf{s}}_2\}$ denote sets grouping the speech sources of interest and their corresponding estimates, respectively. Let $\mathcal{F}$ denote the SS model trained to extract $\hat{S}$ given the input mixture $\mathbf{y}$. $\mathcal{F}$ is trained applying PIT to minimize

\begin{equation}
    \mathcal{L}(S, \hat{S}) = - \frac{1}{2}\max_{\pi}\sum_{n=1}^{2}\mathcal{D}(\mathbf{s}_{\pi(n)}, \hat{\mathbf{s}}_n) \; ,
    \label{eq:pit}
\end{equation}

\noindent where $\pi$ represents the permutation set on $S$ and $\mathcal{D}(\mathbf{s}, \hat{\mathbf{s}})$ is a signal-level similarity measure between a target utterance $\mathbf{s}$ and its estimate $\hat{\mathbf{s}}$. The most commonly used similarity measures are SDR and scale-invariant SDR (SI-SDR) \cite{le2019sdr}. Both measures can be jointly expressed by letting
\begin{equation}\label{eq:SDR}
    \mathcal{D}(\mathbf{s}, \hat{\mathbf{s}}) = 10\log_{10}\left(\frac{\| \alpha \mathbf{s} \|^{2}}{\| \hat{\mathbf{s}} - \alpha \mathbf{s} \|^{2} + \epsilon} + \epsilon\right) \; .
\end{equation} 
where $\|\cdot\|$ denotes Euclidean norm, $\epsilon$ is a constant for numerical stability and $\alpha$ is a parameter selected to be either 1 for SDR or the scalar projection of $\hat{\mathbf{s}}$ onto $\mathbf{s}$, i.e., $\frac{ \hat{\mathbf{s}}^{T}\mathbf{s}}{\| \mathbf{s} \|^{2}}$, for SI-SDR. Among the two measures, SI-SDR is typically preferred due to its invariance to signal scaling. However, both measures are ill-defined when one of the target signals is zero energy, such as $\mathbf{s}_2$ in (\ref{eq:input}). Hence, the problem is to modify the training objective of $\mathcal{F}$ to allow the same model to perform joint SE and SS in single and dual talker scenarios.\par

\section{Existing Solutions}
\label{sec:background}

Let us first discuss previously proposed approaches in the literature for tackling the problem in this study.\par

\subsection{Softmax SDR}
% The signal-level SDR-based similarity measure has been effective for training speech separation models using (\ref{eq:pit}) but may face issues when dealing with zero-energy target signals.

One approach is to add a small positive constant $\epsilon$ to the numerator of the SDR measure. However, this introduces a bias in training since zero-energy target signals are easy to learn. However, this issue can be addressed by limiting the SDR value to a soft maximum \cite{vonneumann21_GraphPIT}, resulting in the following performance measure

% and stabilize the denominator in case of perfect reconstruction

% A small positive constant $\epsilon$ is added to SDR's numerator \cite{vonneumann21_GraphPIT} to define the loss function for zero-energy signals. However, zero-energy signals are easy to learn and can lead to training bias. To address this, a soft maximum is added to limit the SDR value (thresholding) and stabilize the denominator in case of perfect reconstruction \cite{wisdom2020unsupervised}. This modified SDR measure, called $\epsilon-\text{tSDR}$, effectively handles zero-energy target signals and can be defined as follows:

\begin{equation}
    \mathcal{D}_{\epsilon-\text{tSDR}} (\mathbf{s}, \hat{\mathbf{s}}) = 10\log_{10} \frac{\| \mathbf{s}\|^2 + \epsilon }{\| \hat{\mathbf{s}} - \mathbf{s}\|^2 + \tau (\| \mathbf{s}\|^2 + \epsilon)} \; ,
\end{equation}
where $\tau = 10^{-\text{SDR}_{\text{max}}/{10}}$ is a constant that restricts the maximum value of SDR to some threshold $\text{SDR}_{\text{max}}$. 

% It should be noted that modifying the SDR loss function may introduce distortion to speech signals, potentially reducing the effectiveness of separation quality and resulting in an imperfect SDR score.

\subsection{Source aggregated SDR}
% The Source aggregated SDR (SA-SDR), introduced in a recent study \cite{von2022sa}, presents a new method for computing loss in speech separation models. Unlike traditional methods that compute the local SDR for each speaker and then calculate the arithmetic mean as shown in (\ref{eq:pit}), SA-SDR computes a global SDR by aggregating the energies of the target signals in the numerator and the imperfect reconstruction parts from each signal in the denominator. This approach offers a more effective way to compute loss by considering the global energy of the target signals and the reconstruction errors across all speakers. The overall loss function with PIT can be given as,

A recent study proposed a modified training objective for handling varying numbers of overlapping talkers, called source aggregated SDR (SA-SDR) \cite{von2022sa}. Unlike traditional PIT in (\ref{eq:pit}) paired with the SDR measure in (\ref{eq:SDR}), which, for a given permutation, computes the arithmetic mean of signal-level SDRs, SA-SDR aggregates the energies of the target signals and reconstruction errors to compute a global SDR measure. The modified PIT objective is given by
\begin{equation} \label{eq:sa_sdr}
    \mathcal{L}_{\text{SA-SDR}}(S, \hat{S}) = -\max_{\pi}10\log_{10} \frac{\sum_{n=1}^{N} \| \mathbf{s}_{\pi(n)}\|^2}{\sum_{n=1}^{N} \| \hat{\mathbf{s}}_n - \mathbf{s}_{\pi(n)}\|^2} \; ,
\end{equation}
where $N$ is the maximum number of concurrent speakers, which in the context of this work is 2. This training objective is defined as long as the mixture contains at least one active speech source.\par

\subsection{Multi-Objective Loss}
An alternative approach is to introduce distinct loss functions during training for tackling different numbers of talkers \cite{wisdom2021s}. In the context of this work, we can modify the training objective as follows
\begin{equation}
 \mathcal{L}_{\text{MOL}}(S, \hat{S}) = \begin{cases}
-\mathcal{D}(\mathbf{s}_1, \hat{\mathbf{s}}_1) - \lambda \mathcal{D}_{\text{log-MSE}}(\mathbf{s}_2, \hat{\mathbf{s}}_2), & \mathbf{s}_2 = \mathbf{0}\\
\mathcal{L}(S, \hat{S}), & \mathbf{s}_2 \neq \mathbf{0}
\end{cases}
\end{equation}
where $\lambda$ is some positive constant, $\mathbf{0}$ is a vector of zeros, and
\begin{equation} \label{eq:mse}
    \mathcal{D}_{\text{log-MSE}}(\mathbf{s}, \hat{\mathbf{s}}) = -10\log_{10}(\| \hat{\mathbf{s}} - \mathbf{s} \|^2 + \epsilon)
\end{equation}
is the log mean squared error (log-MSE). In a single-talker scenario, this approach employs signal-level SDR/SI-SDR in (\ref{eq:SDR}) along with log-MSE for training the model to respectively output enhanced speech on the first channel and zeros on the second channel. The parameter $\lambda$ is introduced to balance the gradients between the two objectives. In dual-talker scenarios, standard PIT combined with signal-level SDR/SI-SDR is employed.\par

% \footnote{To maintain consistency with the other similarity measures introduced in this paper, $\mathcal{D}_{\text{MSE}}$ returns a more positive value as the similarity between signals increases.}

% Having noted that a dual-talker SS model tends to output a similar signal at both channels when the input mixture consists of only one speech source. 

\section{Proposed Solution} \label{sec:proposed}
The methods discussed so far involve modifying (\ref{eq:pit}) or (\ref{eq:SDR}) to handle zero-energy signals. However, modifying the desired performance measure may degrade the model's performance in terms of the original metric. Hence, in this work we prefer to avoid introducing any modifications to (\ref{eq:pit}) and (\ref{eq:SDR}) and reformulate the target signals in $S$ as follows
\begin{equation} \label{eq:proposed}
\bar{S} = \begin{cases}
\{\mathbf{s}_1, \mathbf{s}_1\}, & \mathbf{s}_2 = \mathbf{0}\\
\{\mathbf{s}_1, \mathbf{s}_2\}, & \mathbf{s}_2 \neq \mathbf{0} \;,
\end{cases}
\end{equation}
meaning that, in single-talker scenarios, $\mathcal{F}$ is trained to extract $\mathbf{s}_1$ at both channels, whereas, in dual-talker scenarios, $\mathcal{F}$ is trained to extract the distinct sources, just as in conventional SS. This approach is motivated by the observation that dual-talker SS models tend to output a similar version of the signal at both channels when the input mixture consists of just one speech source. Hence, the idea is to simplify training without the need of modifying the desired performance measure.\par
% By adjusting the target signals, we can train the speech enhancement and separation model similar to the standard PIT-based source averaged SI-SDR loss commonly used in state-of-the-art speech separation methods. The modified training target loss (MTL) can then be defined as
% \begin{equation}
%     \mathcal{L}_{\text{MTL}}(S, \hat{S}) = \mathcal{L}(\bar{S}, \hat{S}). 
% \end{equation}

\begin{figure*}[t]
\begin{minipage}[b]{0.95\linewidth}  
  \centering
  \centerline{\includegraphics[width=.8\textwidth]{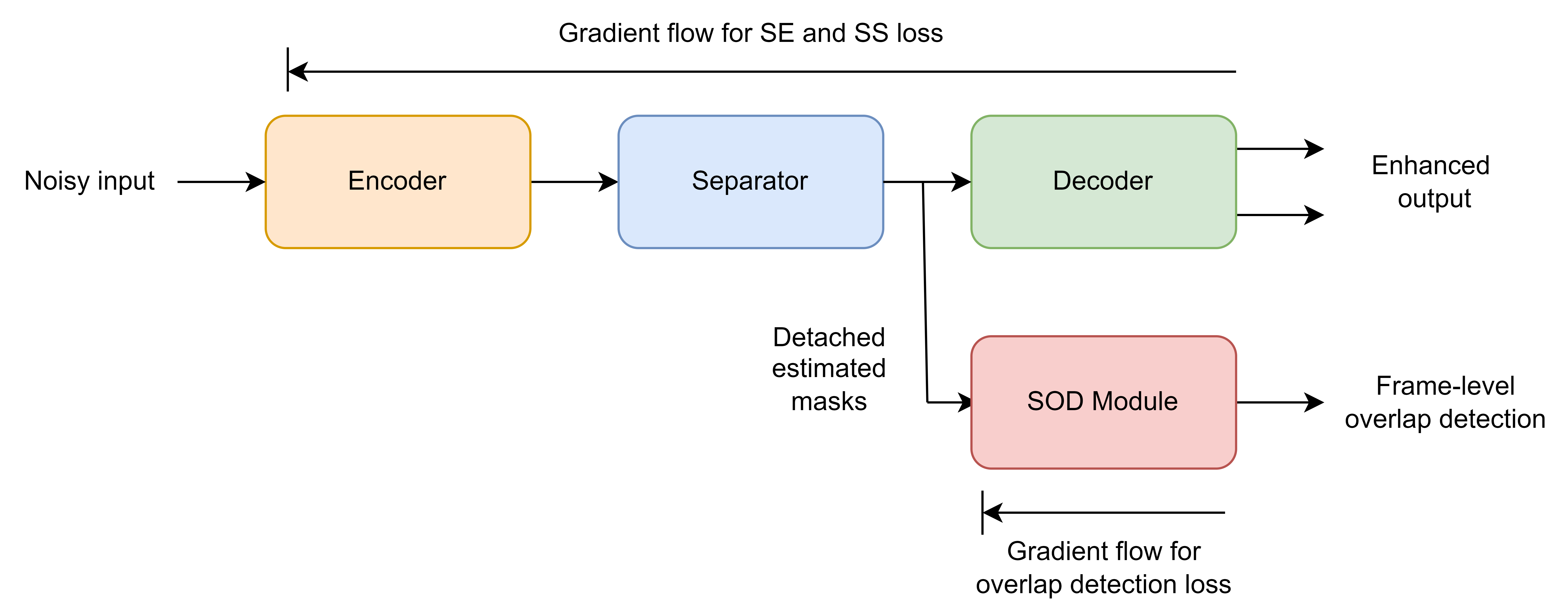}}
\end{minipage}
\caption{Schematic of the proposed methodology consisting of a primary neural network $\mathcal{F}$ for SE and SS tasks with an integrated SOD module. The model $\mathcal{F}$ consists of encoder, separator, and a decoder modules. The arrows indicate the flow of gradients during training.} 
\label{fig:vss}
\end{figure*}

The proposed modification in (\ref{eq:proposed}) introduces the need for an additional post-processor to differentiate between single and dual talker scenarios. For this purpose, we propose a lightweight SOD module that would work in tandem with $\mathcal{F}$ to detect speaker overlap in real time. This module can serve different purposes, e.g., preventing ASR systems from processing both output channels in single-talker scenarios; and, optionally, allowing a means to know when to replace the output at the second channel with zeros if consistency with the original target signals in $S$ is desired.\par

In this work, we consider $\mathcal{F}$ as a real-time time-domain SS neural network that follows the general encoder-separator-decoder design of the well-known time-domain audio speech separation network (TaSNet) \cite{luo2018tasnet}. As shown in Fig. \ref{fig:vss}, the SOD module operates in a frame-wise manner and takes as input the two-channel mask vectors from the separator module of $\mathcal{F}$. The masks are concatenated resulting in the vector $\mathbf{m}^{(2K)}$, where the superscript denotes the length of the vector and $K$ is the dimension of individual-channel masks. $\mathbf{m}^{(2K)}$ is then fed as input to the SOD module, which consists of the following three processing stages
% As an integral aspect of this methodology, we propose to use a speaker overlap detection (SOD) module that harnesses the frame-level masks generated from the primary model $\mathcal{F}$ to detect speaker overlap in real-time. Specifically, consider a conventional time-domain speech separation model incorporating an encoder, separator, and decoder module. By redefining the target signal and utilizing the separator-generated masks as an input feature, we train the SOD module to classify instances of speaker overlap. The proposed approach is illustrated in Figure \ref{fig:vss}.

% Equation (\ref{eq:tiny_ML}) summarizes the operation of the SOD module. 
% The input to the SOD module consists of the concatenated detached masks from the separator module of $\mathcal{F}$, where each channel mask $\mathbf{m}$ contains $K$ features. These concatenated masks, denoted as $\mathbf{m}^{(2K)}$, are then processed in the SOD module, which comprises three layers as follows

\begin{align}
\begin{split}
\mathbf{m}_1^{(H)} &= \text{ReLU}(\text{FF}(\mathbf{m}^{(2K)}))
\\
\mathbf{m}_2^{(H)} &= \text{ReLU}(\text{Stacked-GRU}(\mathbf{m}_1^{(H)}))
\\
\mathbf{m}_3^{(1)} &= \sigma(\text{FF}(\mathbf{m}_2^{(H)})) \;, 
\end{split}
\label{eq:tiny_ML}
\end{align}
where $\text{ReLU}(\cdot)$ and $\sigma(\cdot)$ denote rectified linear unit and sigmoid activation functions, respectively, $\text{FF}(\cdot)$ denotes a feed-forward layer, and $\text{Stacked-GRU}(\cdot)$ denotes two stacked gated recurrent unit (GRU) layers. Recurrent processing is introduced to provide longer context awareness. It follows that the SOD module transforms an input vector of length $2K$ into a lower-dimensional hidden state vector of length $H$ which is then further processed and mapped into a classification output denoted by a scalar value between 0 and 1. Frame-level outputs are averaged across an individual training utterance and the module is trained separately from $\mathcal{F}$ to minimize the binary cross-entropy loss.\par

% For this purpose, the SOD module employs two feed-forward layers (FF), a two stacked gated recurrent unit (GRU) layers, ReLU as a non-linearity layer, and a Sigmoid activation function $\sigma$ to output the classification probabilities. The binary cross-entropy loss function is employed during training, with averaging carried out across all frames. To ensure that the weights of the encoder and separator, which use a signal reconstruction-based loss function, are not impacted by overlap detection, the input masks are detached from the primary speech enhancement and separation model.

Despite its simplicity, the advantage of the proposed training approach in (\ref{eq:proposed}) compared to existing solutions is that it allows training the same model $\mathcal{F}$ on single and dual-talker datasets without modifying the popular SI-SDR-based loss function. Furthermore, forcing both channels to output an audio signal, even in single-talker scenarios, enables the model to equally optimize the parameters associated with the output at both channels. Lastly, the new training approach paired with the proposed SOD post-processing module conveniently enable efficient real-time detection of speaker overlap by reusing the frame-level masks estimated by the separator in $\mathcal{F}$ as input to the SOD block. This claim follows from the reasoning that SOD is simpler when the input consists of the already separated signals instead of the original mixture.\par

\section{Experimental Configurations}
We evaluate the performance of the proposed methodology on SE and SS tasks using an existing neural network architecture.\par
% speech enhancement and separation dataset is generated using publicly available datasets, LibriSpeech \cite{panayotov2015librispeech}, and WHAM \cite{wichern2019wham}.

\subsection{Dataset}
A dataset is generated to simulate single and dual-talker noisy mixtures in a reverberant room. This dataset consists of 36000, 10800, and 9000 4-second long utterances sampled at 16 kHz for training, testing, and validation, respectively. Clean speech and noise utterances are obtained from LibriSpeech \cite{panayotov2015librispeech} and WHAM! \cite{wichern2019wham} datasets, respectively. For each utterance, the room dimensions are randomly sampled between 5 and 10 meters in length and width and 2 to 5 meters in height. The reverberation time is randomly sampled between 0.1 and 0.5 seconds. The number of talkers in the mixture is set to vary from 1 to 2. In dual-talker utterances, speech signals are mixed to have a randomly sampled signal-to-interference ratio between -5 and 5 dB. Speech and noise source positions are randomly sampled within the room with the constraint of being at least 50 cm away from the walls. The microphone is placed at the center of the room, and the image method \cite{habetsroom} is used to generate the corresponding room impulse responses (RIRs). Reverberant speech and noise signals are added and mixed to have a signal-to-noise ratio (SNR) randomly sampled between 5 and 20 dB.\par

\subsection{Network Architecture and Training}
UG-Net \cite{patel2022ux} is adopted as the baseline model for $\mathcal{F}$. UG-Net is a casual TaSNet-like system designed for SS. The dimension of the encoder in UG-Net is set to 256 and the separator depth is set to 5. The frame size is set to 2 ms and 50\% overlap is used, resulting in an algorithmic latency of only 3 ms. The network is trained using the Adam \cite{kingma2014adam} optimizer for 70 epochs with a batch size of 8. The initial learning rate is set to $10^{-3}$ and multiplied by 0.98 every epoch. Gradients are clipped to [-5, 5] during backpropagation to avoid the exploding gradient problem.\par

% As a baseline model for $\mathcal{F}$, we adopted the UG-Net, a causal architecture that employs GRUs \cite{patel2022ux}. This model takes a single channel as an input and generates two output channels. The frame size used is 2 ms, with a 50\% overlap. We set the feature dimension of the encoder module to 256 and employed cumulative layer normalization. The separator module's depth was fixed at 5. The network is trained using the Adam optimizer for 70 epochs with a batch size of 8. The initial learning rate is set to $10^{-3}$ and multiplied by 0.98 every epoch. To prevent the exploding gradient problem, gradients are clipped to [-5, 5] during the backward pass.

Once training of $\mathcal{F}$ was completed, the SOD model was trained using as input the masks estimated by the separation module of UG-Net on the training dataset. The hidden state dimension $H$ of the SOD module was set to 64.\par

% The SOD module for speaker overlap detection was trained simultaneously with the main separation baseline model, employing a similar learning rate and optimizer. However, the training of the overlap detection model was performed independently of the main separation model. This was achieved by detaching the input masks to the overlap-detection module from the baseline model. The hidden state dimension $H$ is set to 64.

\subsection{Evaluation}
The SE and SS performance of the proposed method is evaluated using the following three performance measures: Perceptual Evaluation of Speech Quality (PESQ) \cite{rix2001perceptual}, Short-Time Objective Intelligibility (STOI) \cite{taal2010short}, and SI-SDR in dB. These metrics are reported separately for single and dual-talker scenarios using the formatting PESQ/STOI/SI-SDR. Additionally, we evaluate the SOD module's accuracy in detecting dual-talker segments using frame-level true negative (TNR) and true positive (TPR) rates.\par
% Three performance metrics were employed to evaluate the proposed approach: Perceptual Evaluation of Speech Quality (PESQ) \cite{rix2001perceptual}, Short-Time Objective Intelligibility (STOI) \cite{taal2010short}, and absolute SI-SDR in dB. These metrics are reported as PESQ/STOI/SI-SDR. Two separate evaluations are reported for speech enhancement (SE) and speech separation (SS) subsets of the test set based on the number of active speakers in each utterance, providing a comprehensive understanding of the model's overall performance. We also evaluated the frame-level speaker overlap detection by computing the true negative rate (TNR) and true positive rate (TPR), which measures the model's accuracy in detecting the presence of a second speaker in the audio signal.

\section{Results}
% In the first experiment, the model $\mathcal{F}$ is trained on three different tasks: speech enhancement (SE task), speech separation (SS task), and speech enhancement and separation (SE-SS task). For SE task, the network is trained on a subset of the training dataset that includes only a single active speaker. For SS task, it is trained on a subset of the dataset that includes two active speakers only. For SE-SS task, the whole training dataset is being considered. The model is trained for SE and SS tasks using SDR and SI-SDR loss functions, as the number of active speakers in a mixture is known apriori. For the SE-SS task that deals with zero-energy targets, the model is trained with proposed modified training targets loss and was compared with when trained on soft maximum SDR, SA-SDR and multi-objective loss as discussed earlier in Section \ref{sec:background}. Table \ref{table:loss_table} summarizes the evaluation results. 

Three experiments were conducted. In the first experiment, $\mathcal{F}$ is trained on the following three tasks: SE, SS, and SE-SS. For the SE task, the network is trained on the subset of the training set that includes only single-talker utterances using SDR and SI-SDR measures as training objectives, where only the first output channel is considered. For the SS task, the network is trained on the subset of the training set that includes only dual-talker utterances using the classic PIT with SDR ($\mathcal{L}_\text{SDR}$) and SI-SDR ($\mathcal{L}_\text{SI-SDR}$) training objectives. Finally, for the SE-SS task, the network is trained on the entire training set using the existing training objectives described in Section \ref{sec:background} and the proposed method described in Section \ref{sec:proposed}. The parameters $\epsilon$, $\text{SDR}_{\text{max}}$ and $\lambda$ are set to $10^{-8}$, $30$ dB and $0.1$, respectively. SI-SDR was used as the signal-level similarity measure in $\mathcal{L}_{\text{MOL}}$ and the proposed method.\par

\begin{table}
    \caption{Comparison with existing solutions.}
    \label{table:loss_table}
    \small
    \centering
    \setlength\tabcolsep{1.5pt} % default value: 6pt  
    \begin{tabular}{M{5.2em} M{6.5em}  M{7em} M{7em}}  
            \toprule
            %       \multirow{2}{7em}{\makecell[c]{Training \\ Loss}}            & \multicolumn{2}{c}{SE-SS task} \\
            % \cmidrule(lr){2-3} 
                    Tasks & \makecell[c]{Training \\ Objective}         & Single-talker & Dual-talker \\
              \midrule  
                Unprocessed & -- & 2.36/0.76/7.88 & 1.26/0.49/-0.13 \\
              \midrule
                       \multirow{2}{5.2em}{\makecell[c]{SE}}  & SDR   & \textbf{2.93}/\textbf{0.91}/15.68 &  --  \\ %1.21/0.44/-4.33
                       & SI-SDR   & 2.92/0.90/\textbf{15.79} & -- \\ %1.22/0.43/-6.79
             \midrule
                 \multirow{2}{5.2em}{\makecell[c]{SS}}  & $\mathcal{L}_{\text{SDR}}$   & 2.52/0.77/9.35 & \textbf{1.85}/0.68/\textbf{5.96}  \\
                       & $\mathcal{L}_{\text{SI-SDR}}$   & 2.58/0.78/9.10 & 1.84/\textbf{0.69}/5.89 \\
                \midrule
              \multirow{4}{5.2em}{\makecell[c]{SE-SS}}                 
             & $\mathcal{L}_{\epsilon-\text{tSDR}}$  & 2.80/0.87/13.58 & 1.70/0.65/5.15  \\
             & $\mathcal{L}_{\text{SA-SDR}}$         & \textbf{2.88}/\textbf{0.89}/14.30 & 1.74/0.67/5.49 \\
             & $\mathcal{L}_{\text{MOL}}$            & 2.78/0.88/14.78 & 1.72/0.66/5.19 \\
             & Proposed & 2.87/\textbf{0.89}/\textbf{14.94} & \textbf{1.80}/\textbf{0.68}/\textbf{5.68} \\ 
            \bottomrule 
    \end{tabular}
\end{table}

\begin{figure}[t]
\begin{minipage}[b]{1.0\linewidth}  
  \centering
  \centerline{\includegraphics[width=1\textwidth]{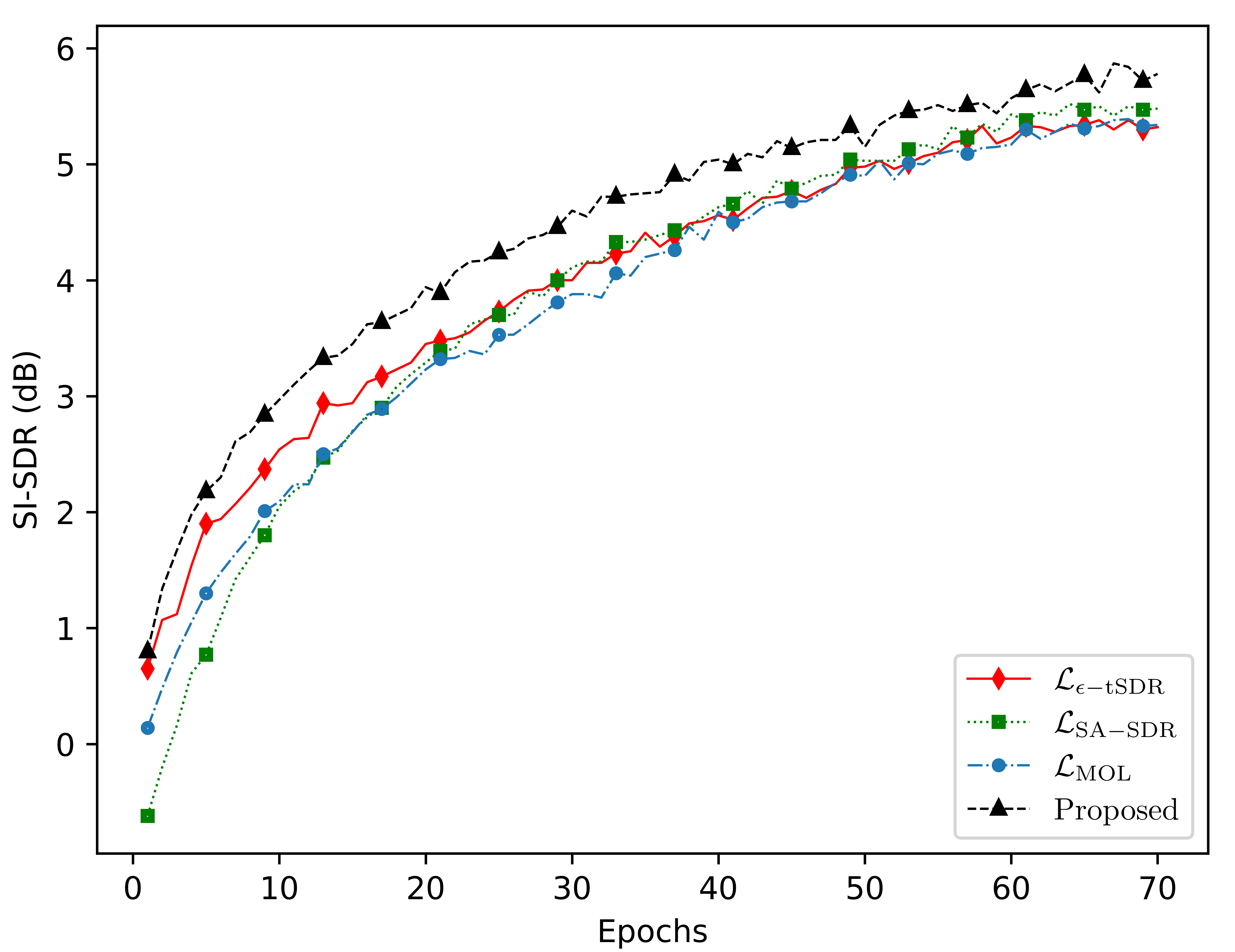}}
\end{minipage}
\caption{Learning curves of the different training objectives targeting the SE-SS task. Results are reported in terms of SS performance on the dual-talker subset of the validation set.}
\label{fig:val}
\end{figure}

Table \ref{table:loss_table} reports the results of the first experiment. We note that training the model solely on the dual-talker set leads to a significant reduction in performance on the single-talker set when compared to all other methods, thus confirming the need for an improved training objective that can handle varying numbers of speakers. Among the training methods suitable for the SE-SS task, the proposed approach is shown to attain the best overall performance, especially in terms of SI-SDR. Fig. \ref{fig:val} further illustrates the SI-SDR performance gap in the learning curves of the different methods. This performance improvement is attributed to the use of an unaltered SI-SDR-based training objective, made possible by the proposed formulation in (\ref{eq:proposed}).\par

% To better understand the learning capabilities of each method for SS in SE-SS tasks, we plotted the SI-SDR score on the validation set as the model was trained, as illustrated in Figure \ref{fig:val}. The proposed method yielded a better learning curve, further substantiating its efficacy.

% \caption{SI-SDR comparison of the proposed approach and different loss functions during training of $\mathcal{F}$ for SE-SS task on SS subset of the validation set.}

In the second experiment, we investigated the effect of various levels of SNR on the performance of the proposed training method on the SE-SS task and the TPR and TNR scores of the proposed SOD module. The results in Table \ref{table:noise} show that TPR tends to be consistently higher than TNR, suggesting that the model finds detecting dual-talker segments easier than single-talker. As expected, we also note that the performance of the proposed methods tends to improve at increased SNR.\par

% In the second experiment, we investigated the effect of various levels of background noise SNR on the performance of the proposed $\mathcal{L}_{\text{MTL}}$ training method and the SOD module for the SE-SS task. The outcomes are presented in Table \ref{table:noise}, which includes the TNR and TPR scores representing the accuracy of the mask-based SOD module in detecting speaker overlap in the input utterance. The findings indicate that the TPR score remained consistently higher than the TNR score, indicating that the model found distinguishing between two speakers in noisy conditions easier than identifying when only one speaker was present. The introduction of background noise significantly impacted both SE and SS performance, leading to decreased quality and perception and reduced accuracy in detecting speaker overlap.\par

\begin{table}
    \label{table:noise}
    \caption{Evaluation of the proposed methodology on the SE-SS task in varying SNR conditions.}
    \small
    \centering
    \setlength\tabcolsep{1.5pt} % default value: 6pt  
    \begin{tabular}{M{5em}  M{3em} M{3em}  M{7em} M{7em}}  
            \toprule
               SNR  & TNR & TPR  & Single-talker & Dual-talker \\
              \midrule                 
             5-10 dB   & 94.6\% & 95.3\% & 2.77/0.86/14.52 & 1.68/0.63/5.13 \\ 
             10-15 dB  & 97.2\% & 97.6\% & 2.90/0.88/15.02 & 1.80/0.68/5.73 \\
             15-20 dB  & 98.7\% & 99.3\% & 2.94/0.90/15.28 & 1.92/0.70/6.10 \\
            \bottomrule 
    \end{tabular}
\end{table}

\begin{table}[h!]
    \caption{Effect of SOD masking on the quality of separated signals.}
    \label{table:mask}
    \small
    \centering
    \setlength\tabcolsep{1.5pt} % default value: 6pt  
    \begin{tabular}{M{4.34em}  M{4.34em} M{6em} M{7em} }  
            \toprule
               TNR   & TPR & SOD masking & Dual-talker \\
              \midrule                 
            \multirow{2}{4.34em}{\makecell[c]{97.9\%}} & \multirow{2}{4.34em}{\makecell[c]{98.6\%}} & \xmark & 1.80/0.68/5.74  \\ 
            & & \cmark & 1.72/0.67/5.53\\
            \bottomrule 
    \end{tabular}
\end{table}

% In the third experiment, we aimed to evaluate the model's ($\mathcal{F}$) reliability when used in conjunction with the SOD module. For this purpose, the second channel output of the model $\mathcal{F}$ was modified based on the classification probability of the mask-based SOD module. Specifically, frames were set to zero if the probability of overlap detection was less than 0.5 and remained unaltered if the probability was greater than 0.5. This approach was tested in a causal frame-based setting. We applied overlap detection masks to the model's output after an initial cooling period of 800 ms to evaluate the effectiveness. The results of this analysis are summarized in Table \ref{table:mask}, where TNR and TPR represent the accuracy of overlap detection after the cooling period. Our observations indicate that the proposed frame-based overlap detection method is reliable and does not significantly impact the overall SS performance when used in conjunction. This suggests we can rely on this approach for tasks such as ASR, where speaker overlap detection is crucial, without sacrificing the system's overall performance.

In the third experiment, we further quantify the detection performance of the SOD module in dual-talker scenarios. For this purpose, we replace the frame-wise output at the second channel of $\mathcal{F}$ with a vector of zeros whenever the frame-level SOD value is below 0.5 and evaluate the resulting extracted signals in terms of PESQ, STOI and SI-SDR. We refer to the procedure of replacing the second channel output frames with zeros when the SOD value is low as \textit{SOD masking}. The initial 500 ms segment of the extracted signals is ignored to warm up the SOD module. Table \ref{table:mask} reports the results. We note that the use of SOD masking does not result in excessive degradation in signal quality, thus confirming the effectiveness of the proposed SOD module.\par

% In the third experiment, we tested the reliability of the model $\mathcal{F}$ when used with the SOD module for speaker overlap detection. To do this, we modified the second channel output of $\mathcal{F}$ based on the classification probability of the SOD module. Specifically, we set frames to zero if the probability of overlap detection was less than 0.5 and left them unchanged if the probability was greater than 0.5. This approach was evaluated in a causal frame-based setting, where we applied overlap detection masks to the model's output after an initial cooling period of 500 ms. The effectiveness of the proposed frame-based overlap detection method was evaluated using TNR and TPR accuracy metrics, which are summarized in Table \ref{table:mask}. Our findings show that the proposed method is reliable and does not significantly impact the SS performance when used in conjunction. This suggests we can use this approach for tasks such as ASR, where speaker overlap detection is critical, without compromising the system's overall performance.

\section{Conclusion}
This paper proposed a simple and practical training approach for leveraging a real-time DL model to perform joint SE and SS in single and dual-talker scenarios. The proposed methodology circumvents the problem of zero-energy target signals by training the separation network to extract speech in both its output channels. The newly defined training targets facilitate the use of the SI-SDR measure during training as in conventional time-domain SS with fixed number of speakers. Additionally, taking advantage of the inherent speaker-counting property of the separation network, an efficient SOD module is introduced for differentiating between single and dual-talker scenarios in real time. Experimental results showed that the proposed training approach outperforms the existing solutions that consist in modifying the loss function to accommodate zero-energy target signals. Finally, the SOD module was shown to attain high performance.\par

\bibliographystyle{IEEEtran}
\bibliography{mybib}

\end{document}